\documentclass[12pt]{article}
\usepackage{longtable}
\newlength{\dummysp}
\newcommand{\beq}{\begin{equation}}
\newcommand{\eeq}{\end{equation}}

\newcommand{\mtxt}[1]{\mathop{\hbox{{\small #1}}}\nolimits}

\newcommand{\half}{{1 \over 2}}

\newcommand{\beqa}{\begin{eqnarray}}
\newcommand{\eeqa}{\end{eqnarray}}
\newcommand{\nnn}{ \nonumber \\ }
\newcommand{\mod}{{\; \mtxt{mod} \; }}
\newcommand{\Zbf}{{{\bf Z}}}
\newcommand{\Tbf}{{{\bf T}}}
\newcommand{\Wbf}{{{\bf W}}}
\newcommand{\Rbf}{{{\bf R}}}
\newcommand{\adot}{{{\dot a}}}
\newcommand{\bdot}{{{\dot b}}}
\newcommand{\ccdot}{{{\dot c}}}
\newcommand{\dddot}{{{\dot d}}}

\newcommand{\gappeq}{\mathrel{\rlap {\raise.5ex\hbox{$>$}}
{\lower.5ex\hbox{$\sim$}}}}
\newcommand{\lappeq}{\mathrel{\rlap{\raise.5ex\hbox{$<$}}
{\lower.5ex\hbox{$\sim$}}}}
\newcommand{\myref}[1]{(\ref{#1})}
\newcommand{\eelat}{{\Lambda_{E_8 \times E_8}}}
\newcommand{\elat}{{\Lambda_{E_8}}}
\newtheorem{thm}{Theorem}
\newtheorem{defn}{Definition}
\newcommand{\Aem}{$ \{ V_A, a_{1A}, a_{3A} \} $}
\newcommand{\Bem}{$ \{ V_B, a_{1B}, a_{3B} \} $}
\newcommand{\opI}{{\bf (I)}}
\newcommand{\opII}{{\bf (II)}}
\begin{document}

\baselineskip=18pt

\hfill LBNL-46839

\hfill UCB-PTH-00/28

\hfill hep-th/0009104

\hfill January 3, 2001

\vspace{35pt}

\begin{center}
{\bf \Large Completion of standard-like embeddings}
\end{center}

\vspace{5pt}

\begin{center}
{\sl Joel Giedt${}^*$}

\end{center}

\vspace{5pt}

\begin{center}

{\it Department of Physics, University of California, \\
and Theoretical Physics Group, 50A-5101, \\
Lawrence Berkeley National Laboratory, Berkeley,
CA 94720 USA.}\footnote{This work was supported in part by the
Director, Office of Science, Office of High Energy and Nuclear
Physics, Division of High Energy Physics of the U.S. Department of
Energy under Contract DE-AC03-76SF00098 and in part by the National
Science Foundation under grant PHY-95-14797.}

\end{center}

\vspace{30pt}

\begin{center}

{\bf Abstract}

\end{center}

\vspace{5pt}

Inequivalent standard-like observable
sector embeddings in $Z_3$ orbifolds with 
two discrete Wilson lines, as determined by 
Casas, Mondragon and Mu\~noz, are completed
by examining all possible ways of embedding
the hidden sector.  The hidden sector
embeddings are relevant to twisted matter
in nontrivial representations of the Standard Model
and to scenarios where supersymmetry breaking is
generated in a hidden sector.  We find a set of 175
models which have a hidden sector gauge
group which is viable for dynamical
supersymmetry breaking.  Only four different
hidden sector gauge groups are possible in these
models.

\vfill

\hbox{}

\newpage

\baselineskip=18pt

\renewcommand{\thetable}{\Roman{table}}

\noindent
One of the distasteful aspects of four-dimensional
heterotic string phenomenology is the glut of
vacua possible in even the most elementary compactification
schemes.  For instance, the lowly $Z_3$ orbifold \cite{orb}
admits an enormously large
number of low energy effective theories,
once nonstandard {\it embeddings}---including
{\it discrete Wilson lines} (described below)---are
allowed.  (The embedding 
dictates how the {\it space group}---the transformation group
used to construct the orbifold---affects
the gauge degrees of freedom in the
underlying string theory.  For a recent
review of heterotic orbifolds, see \cite{blr}.)
However, it was pointed out some time ago
by Casas, Mondragon and Mu\~noz (CMM) that
most of the embeddings are actually redundant, and only
a relatively small set of inequivalent embeddings exist \cite{cmm}.

In heterotic $Z_3$ orbifold models with
discrete Wilson lines, the embedding is expressed in terms
of four sixteen-dimensional vectors:  the {\it twist embedding} $V$
and three Wilson lines $a_1, a_3$ and $a_5$;
each of the four vectors is given
by one-third of a vector belonging to
the $E_8 \times E_8$ root lattice (denoted
here as $\eelat$):
\beq
3V \in \eelat, \qquad 3a_i \in \eelat, \quad \forall
\; i = 1,3,5.
\label{lvc}
\eeq
(In Appendix A we provide a brief review
of the $E_8$ and $E_8 \times E_8$ root
systems, including explicit realizations
of the respective root lattices $\elat$
and $\eelat$.)
It is convenient to denote the vector formed
from the first eight entries of $V$ by $V_A$
and the vector formed from the last eight
entries of $V$ by $V_B$, so that the twist
embedding $V$ may be written as $V = (V_A;V_B)$.
Eq.~\myref{lvc} then implies 
\beq
3 V_A \in \Lambda_{E_8}^{(A)},
\qquad 3V_B \in \Lambda_{E_8}^{(B)},
\label{vab}
\eeq
where $\Lambda_{E_8}^{(A)}$
and $\Lambda_{E_8}^{(B)}$ are the two
copies of the $E_8$ root lattice used
to construct $\eelat$.
Similarly, we write $a_i= (a_{iA};a_{iB})$ for
each $i=1,3,5$.  In addition to
\myref{vab}, constraint \myref{lvc} becomes
\beq
3 a_{iA} \in \Lambda_{E_8}^{(A)},
\qquad 3 a_{iB} \in \Lambda_{E_8}^{(B)},
\qquad \forall \; i = 1,3,5.
\label{aab}
\eeq
The set $\{ V_A, a_{1A}, a_{3A}, a_{5A} \}$ dictates the
space group transformation properties
of the underlying string degrees of
freedom corresponding to the first $E_8$ factor
of the gauge group; i.e, the
set ``embeds the first $E_8$.''  Similarly, the set
$\{ V_B, a_{1B}, a_{3B}, a_{5B} \}$ embeds the
second $E_8$.  For discrete Wilson lines constructions,
the embedding of the gauge degrees of freedom
has the effect of breaking each $E_8$ down to
a rank eight subgroup:
\beq
E_8(A) \to G_O, \qquad E_8(B) \to G_H,
\eeq
where $G_O$ and $G_H$ are usually coined
the ``observable'' and ``hidden''
sector gauge groups.
Typically, $G_O$ and $G_H$ each contain one or
more $U(1)$s, as required to conserve rank.
We note that one must be careful
not to take the terms ``observable''
and ``hidden'' too literally in these models since {\it twisted}
fields ({\it twisted} and {\it untwisted}
refer to choices of closed string boundary
conditions---properties
which also characterize particle states in the
field theory limit)
charged under the nonabelian factors of $G_O$
are typically also charged under $U(1)$s contained
in $G_H$.  
Thus, gauge interactions between observable
and hidden sector fields are generic and
are potentially a worrisome feature
because of experimental constraints
on gauge interactions beyond those of
the Standard Model.  It is also conceivable
that supersymmetry breaking in the hidden
sector may be communicated too forcefully
to the observable sector via these gauge
interactions, even if they are broken at
an intermediate scale.

Models with three generations of
quarks and leptons can be obtained by choosing
the third Wilson line $a_5$ to
vanish, as explained in refs.~\cite{thrgen}.
Consequently, three generation models
of this ilk are specified by the
set of embedding vectors $ \{ V, a_1, a_3 \} $.
For this reason, we will ignore
$a_5$ in the remainder of this article.
The observable sector gauge group $G_O$
is determined entirely by the set of
observable sector embedding vectors
\Aem.
Many such sets lead to a standard-like 
observable sector gauge group $G_O$ of the form
\beq
G_O = SU(3) \times SU(2) \times U(1)^5 .
\label{cmmgo}
\eeq
CMM have determined observable
sector embeddings of this type, with the additional
requirement of quark doublets---$({\bf 3},{\bf 2})$ 
irreducible representations
({\it irreps}) under the $SU(3) \times SU(2)$ subgroup
of \myref{cmmgo}---in the untwisted sector.
It is suprising that CMM have found that any
observable sector embedding satisfying these two
conditions is equivalent to some one of only nine \Aem;
they are displayed in Table \ref{tabcmm}.
Although they argue that these nine observable sector
embeddings are inequivalent, in Appendix B we show that
three more equivalences exist:
\beq
\mtxt{CMM 3} \simeq \mtxt{CMM 1}, \qquad
\mtxt{CMM 5} \simeq \mtxt{CMM 4}, \qquad
\mtxt{CMM 7} \simeq \mtxt{CMM 6}.
\label{cmmiss}
\eeq
Thus, the number of inequivalent observable
sector embeddings satisfying the CMM conditions
is presumably six; we take CMM observable sector embeddings
1, 2, 4, 6, 8 and 9 as representatives of these six.
This does not mean that only six {\it models}
of this type exist.  For each choice of the six inequivalent
\Aem\ there will be many
possible hidden sector embeddings \Bem,
not all of which are equivalent.
CMM have left the hidden sector embedding unspecified
and the purpose of this paper is to enumerate
the allowed ways (up to equivalences) of
embedding the hidden sector.

One might wonder whether or not the hidden
sector embedding has any phenomenological
relevance from the ``low energy'' ($\lappeq$ 100 TeV)
point of view.
We now point out three ways in which
the hidden sector embedding is crucial to understanding
the low energy physics predicted by a given model.
Firstly, the mass-shell conditions for twisted sector
states in the underlying string theory depend on
the full embedding $ \{ V, a_{1}, a_{3} \} $.
It is the solution of the mass-shell conditions
which determines the spectrum of particle states
below the string scale, roughly $10^{17}$~GeV
for the weakly coupled heterotic string.
Thus, the hidden sector embedding
is important because the spectrum of
twisted sector states, including those
charged under the observable
sector gauge group $G_O$, depends on \Bem.
Secondly, it was mentioned above that
twisted sector fields in nontrivial
irreps of $G_O$ are typically charged under $U(1)$
factors contained in the hidden sector gauge
group $G_H$; the spectrum of hidden $U(1)$ charges
will also depend on the hidden sector embedding.
Finally, the hidden sector embedding is
relevant to model building because $G_H$
and the nontrivial matter irreps under
nonabelian factors of $G_H$ play a crucial role in models
of dynamical supersymmetry breaking; for
example, the authors of refs.~\cite{bgw} illustrate
how the mass of the gravitino and supersymmetry breaking
{\it soft terms} are sensitive to the spectrum
and dynamics of the hidden sector.

The allowed ways of completing the embeddings
of Table \ref{tabcmm} may be determined from
the consistency conditions (which ensure
{\it world sheet modular invariance}---a
property which is necessary for the
absence of quantum anomalies---of the
underlying string theory) 
presented in ref.~\cite{thrgen}:
\beq
3 V_B \in \elat, \qquad 3 a_{iB} \in \elat,
\label{vai}
\eeq
\beq
3 V \cdot V \in \Zbf, \qquad
3 a_i \cdot a_j \in \Zbf, \qquad
3 V \cdot a_i \in \Zbf.
\label{pp2}
\eeq
(The consistency conditions \myref{vai} were already
given in \myref{vab} and \myref{aab} above; the
last two equations in \myref{pp2} must hold
for all choices of $i$ and $j$.)
For example, the first embedding in Table \ref{tabcmm}
has $9 V_A \cdot a_{1A} = -2$.  
Then the hidden sector embeddings which complete
CMM 1 must satisfy $9 V_B \cdot a_{1B} = 2 \mod 3$ since
\beq
V \cdot a_1 = V_A \cdot a_{1A} + V_B \cdot a_{1B}
\eeq
and from \myref{pp2} we see that $9 V \cdot a_1$ must
be a multiple of three.

An infinite number of solutions to \myref{vai} and
\myref{pp2} exist, even after the CMM conditions
of \myref{cmmgo} and untwisted $({\bf 3},{\bf 2})$
irreps are imposed.  This does not imply an infinite
number of {\it physically distinct} models.  For
example, trivial permutation redundancies such as
\beq
\pmatrix{ V_B^I \cr a_{1B}^I \cr a_{3B}^I}
\leftrightarrow 
\pmatrix{ V_B^J \cr a_{1B}^J \cr a_{3B}^J},
\qquad \forall \; I,J = 1, \ldots, 8
\label{triv}
\eeq
allow for different embeddings which give identical
physics.  Redundancies related to the signs of
entries also exist (to be addressed later).
Moreover, we will see below that an upper
bound may be placed on the magnitude of the entries
of the embedding vectors; that is, any embedding
with an entry whose magnitude is 
greater than the bound is equivalent to
another embedding which respects the bound.
Once these redundacies are eliminated 
the number of consistent hidden sector 
embeddings is large ($10^4 \sim 10^5$),
though no longer infinite.
However, just as with the observable sector
embeddings, the equivalence relations
pointed out by CMM allow for a dramatic
reduction when one determines the physically
distinct models.  

\begin{table}
\begin{center}
\begin{tabular}{clll}
CMM \# & \hspace{25pt} $3V_A$ & \hspace{25pt} $3a_{1A}$
& \hspace{25pt} $3a_{3A}$ \\ \hline 
1 & (-1,-1,0,0,0,2,0,0) & (1,1,-1,-1,2,0,0,0) & (0,0,0,0,0,0,2,0) \\ 
2 & (-1,-1,0,0,0,2,0,0) & (1,1,-1,-1,-1,-1,0,0) & (0,0,0,0,0,0,2,0) \\
3 & (-1,-1,0,0,0,2,0,0) & (1,1,-1,-1,-1,0,1,0) & (0,0,0,0,0,2,1,1) \\ 
4 & (-1,-1,0,0,0,2,0,0) & (1,1,-1,-1,-1,0,1,0) & (0,0,0,0,0,0,2,0) \\ 
5 & (-1,-1,0,0,0,2,0,0) & (1,1,-1,-1,-1,-1,1,-1) & (0,0,0,0,0,2,1,1) \\
6 & (-1,-1,0,0,0,2,0,0) & (1,1,-1,-1,-1,2,1,0) & (0,0,0,0,0,1,1,2) \\ 
7 & (-1,-1,0,0,0,2,0,0) & (1,1,-1,-1,-1,2,1,0) & (0,0,0,0,0,0,2,0) \\ 
8 & (-1,-1,0,0,0,1,1,0) & (1,1,-1,-1,-1,1,1,1) & (0,0,0,0,0,1,2,1) \\ 
9 & (-1,-1,0,0,0,1,1,0) & (1,1,-1,-1,-1,-2,0,1) & (0,0,0,0,0,0,0,2) \\
\hline
\end{tabular}
\end{center}
\caption{Observable sector embeddings.}
\label{tabcmm}
\end{table}

We have carried out an automated reduction 
using the equivalence relations of CMM,
which they have denoted ``(i)'' through ``(vi)''.
Their operations ``(ii)'' through ``(v)'' would affect
the observable embedding and are thus irrelevant
to our analysis.  This leaves two
equivalence relations,
presented here for ease of reference.
\begin{enumerate}
\item[\opI] The addition of a root lattice
vector $\ell \in \elat$ 
to any one of the vectors 
$V_B, a_{1B}$ or $a_{3B}$;
it is important to stress that any one
of these embedding vectors may be shifted
{\it independently}:
\beq
V_B \to V_B + \ell \qquad \mtxt{or} \qquad
a_{iB} \to a_{iB} + \ell, \quad i = 1 \, \mtxt{or} \, 3.
\label{otI}
\eeq
\item[\opII] A {\it Weyl reflection} performed
{\it simultaneously} on each of the embedding vectors
in the set $\{ V_B, a_{1B}, a_{3B} \}$:
\beq
V_B \to V_B - (V_B \cdot e_j) e_j,
\qquad
a_{iB} \to a_{iB} - (a_{iB} \cdot e_j) e_j,
\quad i = 1 \, \mtxt{and} \, 3.
\label{otII}
\eeq
\end{enumerate}
In keeping with the notation of Appendix A,
$e_j$ is one of the 240 nonzero roots of $E_8$.
In what follows we will refer to these as
operations \opI\ and \opII.

Operation \opI\ corresponds to an invariance
under translations by elements of the
$E_8$ root lattice $\elat$.  This transformation
group is referred to as the {\it lattice group}
associated with $\elat$; we will denote
this group as $\Tbf$.  Since operation \opI\
allows each vector $V_B$, $a_{1B}$ and $a_{3B}$
to be shifted by a different $E_8$ root lattice
vector, it is actually $\Tbf^3 = \Tbf \times \Tbf
\times \Tbf$ which is the corresponding invariance
group.  Operation \opII\
corresponds to an invariance under
the $E_8$ Weyl group, which we denote $\Wbf$.  To systematically
analyze possible equivalences between
different hidden sector embeddings under
operations \opI\ and \opII, it
is therefore vital to have a rudimentary
understanding of these two groups and their
combined action on the representation space
$\Rbf^8$; i.e., real-valued eight-dimensional
vectors such as $V_B, a_{1B}$ and $a_{3B}$.  
It is also helpful to develop a
concise notation for certain essential features
of $\Tbf$ and $\Wbf$.  For these 
purposes we now embark on a minor study
of these two groups.

It is convenient to notate the elements of $\Tbf$
as $T_\ell$, where $\ell$ is the lattice vector
by which the translation is performed:
\beq
T_\ell P = P + \ell, \qquad \ell \in \elat,
\qquad \forall \; P \in \Rbf^8.
\label{trn}
\eeq
Weyl reflections by any of the 240 nonzero $E_8$ roots
belong to $\Wbf$; we write these as $W_i$ with the
subscript corresponding to the $E_8$ root $e_i$
used in the reflection:
\beq
W_i : P \to W_i P = P - (P \cdot e_i ) e_i,
\qquad \forall \; i = 1, \ldots 240 ,
\qquad \forall \; P \in \Rbf^8.
\label{Wdf}
\eeq
It is not difficult to check that for each
of these operators $W_i^2 = 1$, so that each
is its own inverse; thus, the Weyl group
$\Wbf$ can be built up by taking all possible
products of the 240 $W_i$:
\beq
\Wbf = \{1, W_i, W_i W_j, \ldots \}.
\label{Wgd}
\eeq
The $E_8$ Weyl group is a nonabelian
finite group of {\it order} (the number
of elements) $696 \, 729 \, 600$.  On the
other hand, there are only 240 Weyl
reflections $W_i$.  Thus, the generic element
of $\Wbf$ is not a simple reflection \myref{Wdf},
but is a product of several such reflections.
In what follows, we write generic elements
of the Weyl group in calligraphic type:
${\cal W}_I \in \Wbf$, with
$I = 1, \ldots, 696 \, 729 \, 600$.
Thus, for each element ${\cal W}_I$
of $\Wbf$, Weyl reflections $W_j, W_k, \ldots, W_m$
exist such that
\beq
{\cal W}_I = W_j W_k \cdots W_m .
\label{stw}
\eeq
We point out one more property of the Weyl
group $\Wbf$, which we will have occasion to
appeal to below:  an $E_8$ root lattice
vector, when subjected to a Weyl
group transformation,
yields back an $E_8$ root lattice vector.
Explicitly, if $\ell \in \elat$ and
${\cal W}_I \in \Wbf$, then there exists
a $k \in \elat$ such that
\beq
{\cal W}_I \ell = k .
\label{fnk}
\eeq
In mathematical parlance, ${\cal W}_I$ is
an {\it automorphism} of $\elat$.

With these tools in hand, there is a useful theorem
which we can prove.
\begin{thm}
If ${\cal W}_I \in \Wbf$ and $T_\ell \in \Tbf$,
then there exists a $T_k \in \Tbf$
such that ${\cal W}_I T_\ell = T_k {\cal W}_I$.
\end{thm}
To see this, let $P \in \Rbf^8$ and compute
\beq
{\cal W}_I T_\ell P = {\cal W}_I (P + \ell)
= {\cal W}_I P + {\cal W}_I \ell.
\label{lch}
\eeq
The last step follows from the fact that
${\cal W}_I$ is a linear operator---a property which
is evident from \myref{Wdf} and \myref{stw}.
Using \myref{fnk}, the right-handed side of
\myref{lch} can be rewritten
\beq
{\cal W}_I P + {\cal W}_I \ell = {\cal W}_I P + k = T_k {\cal W}_I P .
\eeq
I.e., ${\cal W}_I T_\ell = T_k {\cal W}_I$, as was to be
shown.

A sequence of operations \opI\ and \opII\
has the form of a product of various elements of
$\Tbf$ and $\Wbf$.  Theorem 1 allows one to rewrite
any sequence of operations \opI\ and
\opII, whatever the order and number of
operations of each type, in the form
\beq
{\cal O} = T_\ell {\cal W}_I,
\qquad T_\ell \in \Tbf,
\qquad {\cal W}_I \in \Wbf .
\eeq
We stress that the element $T_\ell$ may be different
for each of the embedding vectors $V_B$, $a_{1B}$
and $a_{3B}$, but that the Weyl group element ${\cal W}_I$
acting on these vectors must be the same.
Typically, ${\cal W}_I$ will
be a generic element of the Weyl group
taking the form \myref{stw},
corresponding to a string of operations of type
\opII.  Thus, we arrive at the following rather useful
conclusion:  any sequence of operations
\opI\ and \opII, whatever the order
and number of operations of each type, is equal
in effect to a sequence of operations
of type \opII, followed by a {\it single}
operation of type \opI, allowing for
different shifts for each of the three
embedding vectors.  Symbolically,
we need only consider equivalences of the form
\beq
{\cal O} = T_\ell W_j W_k \cdots W_m .
\eeq

Suppose two embeddings $\{ V_B, a_{1B}, a_{3B} \}$
and $\{ V_B', a_{1B}', a_{3B}' \}$.  We want to
determine whether these two embeddings are 
equivalent.  Based on the results of the last
paragraph, we see that it is sufficient to
first tabulate all points in the {\it orbit}
of $\{ V_B, a_{1B}, a_{3B} \}$
under $\Wbf$, and then to check whether any of these
points are related to $\{ V_B', a_{1B}', a_{3B}' \}$
by operation \opI.
(The orbit of $\{ V_B, a_{1B}, a_{3B} \}$
under $\Wbf$ is tabulated by computing the
transformations
$\{ {\cal W}_I V_B, {\cal W}_I a_{1B}, {\cal W}_I a_{3B} \}$
for all $696 \, 729 \, 600$ elements ${\cal W}_I$ of
the $E_8$ Weyl group.)  If the two embeddings
are related in this way, then they are equivalent.

As mentioned above, for a given \Aem, the number
of consistent \Bem\ is infinite; the following
definition exploits operation \opI\ to immediately
and efficiently eliminate enough redundancy
to obtain a finite set.
\begin{defn}
An embedding $\{ V_B, a_{1B}, a_{3B} \}$ is in
{\bf minimal} form provided:
\begin{enumerate}
\item[{\rm (a)}] $3 V_B^I \in \Zbf$, $3 a_{1B}^I \in \Zbf$
and $3 a_{3B}^I \in \Zbf$ for each choice $I = 1, \ldots, 8$;
\item[{\rm (b)}] $|3 V_B^I| \leq 2$, $|3 a_{1B}^I| \leq 2$ 
and $|3 a_{3B}^I| \leq 2$ for each choice $I = 1, \ldots, 8$;
\item[{\rm (c)}] no more than one entry of each vector
$3V_B$, $3 a_{1B}$ and $3 a_{3B}$ has absolute value two,
and any such entry is the left-most nonzero entry.
\end{enumerate}
\end{defn}
Any embedding may be reduced to minimal form by means of
operation \opI.  We will demonstrate the veracity
of this statement by considering $V_B$ which are
not minimal.  It will be understood that similar
statements hold for $a_{1B}$ and $a_{3B}$ which
are not minimal, since operations of type \opI\
are allowed to act independently on $V_B$, $a_{1B}$ and
$a_{3B}$.

From \myref{vai} one sees that $3 V_B$ is an $E_8$
root lattice vector.  As explained in Appendix A,
the entries of an $E_8$ root lattice vector are either
all integral or all half-integral.  In the latter case,
part (a) of Definition 1 will not be satisfied.  However,
operation \opI\ allows us to shift
\beq
3 V_B \to 3 V_B + 3 \ell,
\qquad \ell \in \elat .
\label{13i}
\eeq
If we take $\ell$ to be any lattice vector with half-integral
entries, then \myref{13i} transforms $3 V_B$ to a lattice
vector with integral entries.
Now suppose $3 V_B$ satisfies part (a) of Definition 1
but $| 3 V_B^I| > 2$ for one or more choices of $I$.
It is in all cases possible to find a lattice
vector $\ell$ such that \myref{13i} generates an
equivalent $3 V_B$ which satisfies part (b)
of Definition 1.  To see this, first note
that repeated shifts \myref{13i} by vectors
\beq
3 \ell \in \left\{ \pm (\underline{ 3, 3, 0,0,0,0,0,0}), \;
(\underline{3, -3, 0,0,0,0,0,0})
\right\}
\label{13iii}
\eeq
(underlining indicates that any permutation
of entries may be taken)
allows $3 V_B$ to be translated to a form
where no entry has absolute value greater
than three.  If the original $3 V_B$ satisfied
\myref{vai}, then the translated one will as
well, since the sum of two lattice vectors
is also a lattice vector.  As explained in Appendix A,
an $E_8$ root lattice vector must have its
entries sum to an even number (the final
condition in \myref{eld}).  Then from \myref{vai}
we know that
\beq
\sum_{I=1}^8 3 V_B^I = 0 \mod 2 .
\label{13ii}
\eeq
If for any $I$ the translated vector
has $3 V_B^I = \pm 3$, then \myref{13ii}
implies that there must be a $J \not= I$ such
that $3 V_B^J$ is an odd integer.  If
$3 V_B^J = \pm 3$, then a final shift by
one of the vectors in \myref{13iii} allows
us to set $V_B^I \to 0$ and $V_B^J \to 0$.
For example:
$$
3 V_B = (\ldots,3,\ldots,3,\ldots) \qquad \mtxt{and} \qquad
3 \ell = (\ldots,-3,\ldots,-3,\ldots)
$$
\beq
\mtxt{gives} \qquad
3 V_B \to 3 V_B + 3 \ell = (\ldots,0,\ldots,0,\ldots).
\eeq
On the other hand, if
$3 V_B^J = \pm 1$, then a final shift by
one of the vectors in \myref{13iii} allows
us to set $V_B^I \to 0$ and $V_B^J \to \mp 2$.
From the above manipulations, it should be clear
that a shift \myref{13i} by an appropriate vector
\myref{13iii} will eliminate any pair of $\pm 2$s
appearing in $3 V_B$ in favor of a pair of $\pm 1$s.
Similarly, if a $\pm 1$ precedes a $\pm 2$ (reading
left to right), the order may be reversed---possibly
altering signs---by a shift \myref{13i} by an
appropriate vector \myref{13iii}.
In this way, we are always able to transform any
$V_B$ satisfying parts (a) and (b) of Definition 1
into an equivalent form which also satisfies part (c)
of Definition 1.

It is a simple excercise to verify that
Weyl reflections \myref{Wdf} using $E_8$ roots of the form
$e_i = (\underline{1,-1,0,\ldots,0})$ exchange
two entries; it is also easy to check that
Weyl reflections using roots of the form
$e_i = (\underline{1,1,0,\ldots,0})$
exchange two entries and flip both signs.
We will refer to these as ``integral'' Weyl
reflections.  The second type uses
$E_8$ roots of the form $e_i = (\pm 1/2, \ldots, \pm 1/2)$
with an even number of positive entries,
and we will refer to these as ``half-integral''
Weyl reflections.  These tend to have more
dramatic effects; for example, $3 V_B = (1,\ldots,1)$
can be reflected to $3 V_B = (2,2,0,\ldots,0)$
using $e_i = (1/2,1/2,-1/2,\ldots,-1/2)$.
By such manipulations, together with operation \opI,
it is well-known that only five inequivalent
twist embeddings $V=(V_A;V_B)$ exist (including $V=0$).
Consistency with a given CMM $V_A$
restricts $V_B$ to one or two
choices.  We can eliminate 
remaining redundancies related
to integral Weyl reflections by enforcing
ordering and sign conventions on $a_{1B}$ and $a_{3B}$.
With this in mind, we make the following definition.
\begin{defn}
An embedding $\{ V_B, a_{1B}, a_{3B} \}$
is in {\bf canonical} form if
$3 V_B = (2,1,1,0,0,0,0,0)$ for CMM 1 $\sim$ 7,
$3 V_B = (1,1,0,0,0,0,0,0)$ or $3 V_B
= (2,1,1,1,1,0,0,0)$ for CMM 8 $\sim$ 9; and,
$a_{1B}$ and $a_{3B}$ are {\bf first} fixed to minimal form,
and {\bf then} subjected to whatever integral Weyl reflections
are required such that they satisfy the following conditions:
\begin{enumerate}
\item[{\rm (a)}]
	$V_B^I = V_B^{I+1} \; \Rightarrow \; a_{1B}^I \geq a_{1B}^{I+1}, \;
	I=1,\ldots,7$;
\item[{\rm (b)}]
	$V_B^I=0 \; \Rightarrow \; a_{1B}^I \geq 0, \; I = 3,\ldots,7$;
\item[{\rm (c)}]
	$a_{1B}^7 = 0 \; \Rightarrow \; a_{1B}^8 \geq 0$ while
	$a_{1B}^7 \not= 0 \; \Rightarrow \; 3 a_{1B}^8 \geq -1$;
\item[{\rm (d)}]
	$V_B^I = V_B^{I+1}$ and $a_{1B}^I = a_{1B}^{I+1} \;
	\Rightarrow \; a_{3B}^I \geq a_{3B}^{I+1}, \;
	I=1,\ldots,7$;
\item[{\rm (e)}]
   $V_B^I=a_{1B}^I=0 \; \Rightarrow \;
   a_{3B}^I \geq 0,\; I = 3,\ldots,6$;
\item[{\rm (f)}]
   $a_{1B}^7 = a_{1B}^8 =0$ or
   $a_{1B}^6 = a_{1B}^7 = a_{3B}^6 = 0 \; \Rightarrow \; a_{3B}^7 \geq 0$;
\item[{\rm (g)}]
   $a_{1B}^6 = a_{1B}^7 = 0$ and $a_{3B}^6 \not= 0$ and 
	$a_{1B}^8 \not= 0 \; \Rightarrow \; 3 a_{3B}^7 \geq -1$;
\item[{\rm (h)}]
	$a_{1B}^7 = a_{1B}^8 = a_{3B}^7 = 0 \; \Rightarrow \;
	a_{3B}^8 \geq 0$;
\item[{\rm (i)}]
	$a_{1B}^7 = a_{1B}^8 =0$ and $a_{3B}^7 \not= 0
	\; \Rightarrow \; 3 a_{3B}^8 \geq -1$;
\end{enumerate}
\end{defn}
It is straightforward, though tedious, to
verify that any $a_{1B}$ and $a_{3B}$ of minimal
form can be transformed to satisfy the conditions
listed above using the integral Weyl reflections;
we do not present a proof here as the manipulations
are lengthy and elementary.
Transforming all embeddings $\{ V_B, a_{1B}, a_{3B} \}$
to canonical form, we arrive at a set for which
no two are related purely by integral Weyl reflections.

With the definition \myref{Wdf},
it is not difficult to check
\beq
W_i W_j W_i = W_k, \qquad
e_k = e_j - (e_j \cdot e_i) e_i.
\label{trf}
\eeq
Recall that the entries of $E_8$ roots $e_i$ 
are either all integral or all half-integral.
We denote integral roots with undotted subscripts from the
beginning of the alphabet, $e_a, e_b, \ldots$ and
half-integral roots with dotted subscripts from
the beginning of the alphabet, $e_\adot, e_\bdot, \ldots$.
It should be clear that $e_\adot - (e_\adot \cdot e_a) e_a$
is a half-integral root since $e_\adot \cdot e_a \in \Zbf$.
Thus we can specialize \myref{trf} to obtain, for example,
\beq
W_a W_\adot W_a = W_\ccdot, \qquad
e_\ccdot = e_\adot - (e_\adot \cdot e_a) e_a.
\label{poi}
\eeq
We can then perform manipulations such as
\beq
W_\adot W_a = W_a W_a W_\adot W_a = W_a W_\ccdot,
\label{pf1}
\eeq
\beq
W_\adot W_\bdot W_a = W_a W_a W_\adot W_a W_a W_\bdot W_a
= W_a W_\ccdot W_\dddot,
\label{pf2}
\eeq
where $W_\ccdot$ is defined explicitly in \myref{poi} and
$W_\dddot = W_a W_\bdot W_a$ is defined analogously.
This illustrates how \myref{poi} allows
us to write a generic element \myref{stw} of the Weyl
group $\Wbf$ in the form
\beq
{\cal W}_I = W_a \cdots W_c W_\adot \cdots W_\ccdot .
\eeq
Equivalences related to the string of 
integral Weyl reflections $W_a \cdots W_c$
are eliminated by going to canonical form.
From these considerations
we find that, given a set of canonical embeddings,
equivalences may be identified by the following
procedure:  
\begin{enumerate}
\item[(i)] compute the orbit of 
\Bem\ under strings of half-integral
Weyl reflections;
\item[(ii)] fix the results of (i) to minimal form
by operations of type \opI;
\item[(iii)] fix the results of (ii) to canonical form by
integral Weyl reflections; 
\item[(iv)] check
whether the results of (iii) 
are related by operation \opI\ to any
other embedding in the original set.
\end{enumerate}
The last
step is simply a matter of checking whether
the differences 
$V_B - V_B'$, $a_{1B}-a_{1B}'$
and $a_{3B}-a_{3B}'$ each give
lattice vectors, where \Bem\ is a result
of step (iii) and $\{ V_B', a_{1B}', a_{3B}' \}$
is an element of the original set of
canonical embeddings.

In our automated analysis, we first generated
a list of all possible consistent embeddings
of the hidden sector, constraining them to
be of canonical form.  Since all embeddings can
be reduced to canonical form by way of operations
\opI\ and \opII, we are assured that this list is complete.
The number of ``initial'' embeddings was
at this point already reduced to roughly $10^4$.
Using the procedure outlined in the
previous paragraph, we removed as many of the
redundant embeddings as performing
only 1, 2 and 3 half-integral Weyl reflections
in step (i) would allow.
Because the $E_8$ Weyl group is so large, it proved
to be impractical to act on the initial
embeddings with each of its elements.  It also
proved impractical to perform four or more
half-integral Weyl reflections.  The number
of positive half-integral roots is 64
(negative roots generate the same
Weyl reflections); four Weyl reflections would have
required roughly $10^7$ different operations
for each embedding.
The initial list was thereby
reduced to a mere 192 embeddings.  
This list is guaranteed
to be complete, but entries of the list
are not necessarily inequivalent.
However, already in
going from 2 half-integral Weyl reflections
to 3 half-integral Weyl reflections, the
list did not shrink by much.  It would
appear that though there may be some
equivalences remaining, there should not
be very many.
(It is worth pointing out that application
of an analogous procedure to the observable
sector embeddings turned up equivalences
overlooked by CMM, already at the level
of one half-integral Weyl reflection.)

We have, in addition, determined the hidden
sector gauge group $G_H$ for each of the 192
embeddings.  Only five $G_H$ were found to
be possible, displayed in Table \ref{ghcase}.
This is remarkable, considering that
one might naively expect a large subset of the
112 breakings \cite{hm} of $E_8$ to be present.
Apparently, the CMM requirements of \myref{cmmgo} 
and untwisted quark doublets significantly 
affect what is possible in the hidden sector.

\begin{table}
\begin{center}
\begin{tabular}{cc}
Case & $G_H$ \\ \hline
1 & $SO(10) \times U(1)^3$ \\
2 & $SU(5) \times SU(2) \times U(1)^3$ \\
3 & $SU(4) \times SU(2)^2 \times U(1)^3$ \\
4 & $SU(3) \times SU(2)^2 \times U(1)^4$ \\
5 & $SU(2)^2 \times U(1)^6$ \\
\hline
\end{tabular}
\end{center}
\caption{Allowed hidden sector gauge groups $G_H$.}
\label{ghcase}
\end{table}

In Appendix C, we present lists of
the hidden sector embeddings which complete the CMM
analysis.  We have not displayed Case 5 $G_H$ models,
since we do not regard them as affording viable scenarios
of hidden sector dynamical supersymmetry breaking.
They are, however, available from the author upon
request.  Eliminating the Case 5 $G_H$
models from the total of 192, we are left
with 175 models.  Also not included is the enumeration
of the spectrum of massless matter for these models,
with their $U(1)$ charges.  We have performed
this analysis and hope to present interesting
examples and a summary of general features 
in a later publication.

\vspace{20pt}

\noindent
{\bf \Large Acknowledgements}

\vspace{5pt}

The author would like to thank Prof.~Mary K.~Gaillard
for encouragement during the development of the work 
contained here.  Thanks are also due to
Luis Ib\'a\~nez for helpful communications
at early stages in this research.
This work was supported in part by 
the Director, Office of Science, Office of High Energy and Nuclear
Physics, Division of High Energy Physics of the U.S. Department of
Energy under Contract DE-AC03-76SF00098 and in part by the National
Science Foundation under grant PHY-95-14797.

\vspace{20pt}

\noindent
{\bf \Large Appendix A}

\vspace{5pt}

\noindent
In this appendix we briefly review some salient
aspects of the $E_8$ and $E_8 \times E_8$
root systems.  The material given below can be found in
standard textbooks on string theory,
such as \cite{gsw}, as well as texts on
Lie algebras and groups, such as \cite{corn}; 
it is included here
for ease of reference.

A basis in the root space may be chosen such
that the $E_8$ root lattice can be written
as the (infinite) set of eight-dimensional
vectors
\beq
\elat  =  \left\{ (n_1, \ldots, n_8), \; (n_1 + \half, \ldots, n_8 + \half)
\quad \left| \quad n_1, \ldots, n_8 \in \Zbf, \; \;
\sum_{i=1}^8 n_i = 0 \mod 2 \right. \right\}.
\label{eld}
\eeq
Note that the components of a given $E_8$ root lattice
vector are either all integral or all half-integral.
Lattice vectors $\ell \in \elat$ which satisfy
$\ell \cdot \ell = 2$ (where the ordinary eight-dimensional
``dot product'' is implied) yield the 240
nonzero $E_8$ roots, which we denote $e_1, \ldots, e_{240}$.
By convention, we take as {\it positive roots} those $e_i$
whose first nonzero entry (counting left to right) is
positive.  A {\it simple root} is a positive root which
cannot be obtained from the sum of two positive roots.
Eight simple roots exist for $E_8$, which we denote by
$\alpha_1, \ldots, \alpha_8$.  These form a basis for
the $E_8$ root lattice given in \myref{eld}, which may
alternatively be written as
\beq
\elat = \left\{ \left. \sum_{i=1}^8 m_i \alpha_i \quad
\right| \quad m_1, \ldots, m_8 \in \Zbf \right\}.
\eeq
The $E_8 \times E_8$ root lattice is constructed
by taking the direct sum of two copies of $\elat$,
which we distinguish by labels $(A)$ and $(B)$:
\beq
\eelat = \Lambda_{E_8}^{(A)} \oplus \Lambda_{E_8}^{(B)}.
\eeq
Thus, an $E_8 \times E_8$ root lattice vector $\ell$ is
a sixteen-dimensional vector satisfying
\beq
\ell = ( \ell_A ; \ell_B ), \qquad
\ell_A \in \Lambda_{E_8}^{(A)}, \qquad 
\ell_B \in \Lambda_{E_8}^{(B)},
\eeq
where we have denoted the first eight entries of $\ell$
by $\ell_A$ and the last eight entries of $\ell$
by $\ell_B$, as in the main text.  The 480 nonzero
roots of $E_8 \times E_8$ are given in this
notation by $(e_i;0)$ and $(0;e_i)$, where $e_i$
is one of the 240 nonzero $E_8$ roots.  Similarly,
the sixteen simple roots of $E_8 \times E_8$ are given
by $(\alpha_i;0)$ and $(0;\alpha_i)$, where
$\alpha_i$ is one of the eight $E_8$ simple roots.

\vspace{20pt}

\noindent
{\bf \Large Appendix B}
\vspace{5pt}

\noindent
The equivalences \myref{cmmiss} were uncovered using the
automated routines developed for the analysis of hidden
sector embeddings; any further equivalences
between the observable sector embeddings of CMM would require
four or more half-integral Weyl reflections, transformations
which were not studied for reasons explained
above.  Because the equivalences
\myref{cmmiss} are a significant revision to the results
of ref.~\cite{cmm}, we have chosen to explicitly
demonstrate them in this appendix.
In addition to operations \opI\ and \opII\ used in the main text,
we make use of two redefinitions of the Wilson lines which give
equivalent embeddings (cf.~ref.~\cite{cmm}):
\beq
a_1 \to a_1' = -a_1 - a_3, \qquad
a_3 \to a_3' = a_1 - a_3 ;
\label{B1}
\eeq
\beq
a_1 \to a_1' = a_1 - a_3, \qquad
a_3 \to a_3' = a_1 + a_3 .
\label{B2}
\eeq
In what follows we will ignore the hidden sector embedding
vectors, since in the end we complete the observable
sector embeddings with all consistent choices.

First consider CMM 3, as given in Table \ref{tabcmm}.  We Weyl
reflect (operation \opII) by
$e=\half (1,1,-1,-1,-1,1,1,-1)$ to obtain
\beqa
3V_A & \to & 3V_A' = (-1,-1,0,0,0,2,0,0), \nnn
3a_{1A} & \to & 3a_{1A}' = \half (-1,-1,1,1,1,-3,-1,3), \nnn
3a_{3A} & \to & 3a_{3A}' = \half (-1,-1,1,1,1,3,1,3).
\label{B3}
\eeqa
Application of \myref{B1} yields
\beqa
3a_{1A}' & \to & 3a_{1A}'' = -3a_{1A}' - 3a_{3A}'
= (1,1,-1,-1,-1,0,0,-3), \nnn
3a_{3A}' & \to & 3a_{3A}'' = 3a_{1A}' - 3a_{3A}'
= (0,0,0,0,0,-3,-1,0).
\eeqa
Finally, we employ operation \opI\ to shift
\beq
3a_{1A}'' \to 3a_{1A}''' = 3a_{1A}'' + 3 \ell_1,
\qquad
3a_{3A}'' \to 3a_{3A}''' = 3a_{3A}'' + 3 \ell_3,
\label{B5}
\eeq
where
\beq
3 \ell_1 = (0,0,0,0,3,0,0,3), \qquad
3 \ell_3 = (0,0,0,0,0,3,3,0),
\eeq
to obtain
\beq
3 a_{1A}''' = (1,1,-1,-1,2,0,0,0), \qquad
3 a_{3A}''' = (0,0,0,0,0,0,2,0).
\eeq
With $V_A'$ as given in \myref{B3}, one can see by
comparison to Table \ref{tabcmm} that
$\{ V_A', a_{1A}''', a_{3A}''' \}$ is precisely
the observable sector embedding of CMM 1;
thus, we have shown the first equivalence of
\myref{cmmiss}.

Next consider CMM 5.  We Weyl reflect by
$e = \half (1,1,-1,-1,-1,1,-1,1)$ to obtain
\beqa
3V_A & \to & 3V_A' = (-1,-1,0,0,0,2,0,0), \nnn
3a_{1A} & \to & 3a_{1A}' = \half (1,1,-1,-1,-1,-3,3,-3), \nnn
3a_{3A} & \to & 3a_{3A}' = \half (-1,-1,1,1,1,3,3,1).
\eeqa
Application of \myref{B2} yields
\beqa
3a_{1A}' & \to & 3a_{1A}'' = 3a_{1A}' - 3a_{3A}'
= (1,1,-1,-1,-1,-3,0,-2), \nnn
3a_{3A}' & \to & 3a_{3A}'' = 3a_{1A}' + 3a_{3A}'
= (0,0,0,0,0,0,3,-1).
\eeqa
Shifting as in \myref{B5}, but with
\beq
3 \ell_1 = (0,0,0,0,0,3,0,3), \qquad
3 \ell_3 = (0,0,0,0,0,0,-3,3),
\eeq
we obtain
\beq
3 a_{1A}''' = (1,1,-1,-1,-1,0,0,1), \qquad
3 a_{3A}''' = (0,0,0,0,0,0,0,2).
\eeq
Performing a Weyl reflection of
$\{ V_A', a_{1A}''', a_{3A}''' \}$ by
the root $e' = (0,0,0,0,0,0,1,-1)$ interchanges
entries seven and eight of each embedding
vector:
\beqa
3V_A' & \to & 3V_A'' = (-1,-1,0,0,0,2,0,0), \nnn
3a_{1A}''' & \to & 3a_{1A}'''' = (1,1,-1,-1,-1,0,1,0), \nnn
3a_{3A}''' & \to & 3a_{3A}'''' = (0,0,0,0,0,0,2,0).
\eeqa
Comparing to Table \ref{tabcmm}, we see that
$\{ V_A'', a_{1A}'''', a_{3A}'''' \}$ is the
observable sector embedding of CMM~4; this
proves the second equivalence of \myref{cmmiss}.

Finally consider CMM 7.  Weyl reflection by
$e = \half (1,1,-1,-1,-1,1,1,-1)$ yields
\beqa
3V_A & \to & 3V_A' = (-1,-1,0,0,0,2,0,0), \nnn
3a_{1A} & \to & 3a_{1A}' = (-1,-1,1,1,1,0,-1,2), \nnn
3a_{3A} & \to & 3a_{3A}' = \half (-1,-1,1,1,1,-1,3,1).
\eeqa
Application of \myref{B2} gives
\beqa
3a_{1A}' & \to & 3a_{1A}'' = 3a_{1A}' - 3a_{3A}'
= \half (-1,-1,1,1,1,1,-5,3), \nnn
3a_{3A}' & \to & 3a_{3A}'' = 3a_{1A}' + 3a_{3A}'
= \half (-3,-3,3,3,3,-1,1,5).
\eeqa
We shift as in \myref{B5}, but with
\beq
3 \ell_1 = \half (3,3,-3,-3,-3,3,3,-3), \qquad
3 \ell_3 = \half (3,3,-3,-3,-3,3,-3,-9),
\eeq
to obtain
\beq
3 a_{1A}''' = (1,1,-1,-1,-1,2,-1,0), \qquad
3 a_{3A}''' = (0,0,0,0,0,1,-1,-2).
\eeq
Weyl reflection of
$\{ V_A', a_{1A}''', a_{3A}''' \}$ by
$e' = (0,0,0,0,0,0,1,-1)$ then
$e'' = (0,0,0,0,0,0,1,1)$ flips the signs of
entries seven and eight of each embedding
vector, yielding
\beqa
3V_A' & \to & 3V_A'' = (-1,-1,0,0,0,2,0,0), \nnn
3a_{1A}''' & \to & 3a_{1A}'''' = (1,1,-1,-1,-1,2,1,0), \nnn
3a_{3A}''' & \to & 3a_{3A}'''' = (0,0,0,0,0,1,1,2).
\eeqa
Comparing to Table \ref{tabcmm}, we see that
$\{ V_A'', a_{1A}'''', a_{3A}'''' \}$ is the
observable sector embedding of CMM~6; this
demonstrates the third equivalence of \myref{cmmiss}.

\vspace{20pt}

\noindent
{\bf \Large Appendix C}
\vspace{5pt}

\noindent
To construct the full sixteen-dimensional
embedding vectors $V,a_1$, and $a_3$,
simply take the direct
sum of a CMM observable sector embedding (labeled by
subscript $A$) and a hidden sector embedding (labeled
by subscript $B$) from a corresponding table:
\beq
V = (V_A;V_B), \qquad
a_1 = (a_{1A};a_{1B}), \qquad
a_3 = (a_{3A};a_{3B}).
\eeq
For instance, the observable sector embedding CMM~1
from Table~\ref{tabcmm} may be completed by any of the
embeddings in Table~\ref{tab1}.  Any other hidden
sector embedding which is consistent with CMM~1
will be equivalent to one of the choices given in
Table~\ref{tab1}.
It should be noted that CMM~8 and CMM~9 each allow
two inequivalent hidden sector twist embeddings
$V_B$; as a consequence, two hidden sector embedding
tables are given for each.  We have abbreviated
$G_H$ by the cases defined in Table \ref{ghcase}.

{ \footnotesize
\begin{longtable}{|r|l|l|r||r|l|l|r|}
\caption{CMM 1, $3V_{B}$ = (2,1,1,0,0,0,0,0).\label{tab1}} \\
\hline
\# & \hspace{25pt} $3a_{1B}$ & \hspace{25pt} $3a_{3B}$ & $G_H$ &
\# & \hspace{25pt} $3a_{1B}$ & \hspace{25pt} $3a_{3B}$ & $G_H$ \\
\hline \hline \endfirsthead
\caption{(continued) CMM 1, $3V_{B}$ = (2,1,1,0,0,0,0,0).} \\
\hline
\# & \hspace{25pt} $3a_{1B}$ & \hspace{25pt} $3a_{3B}$ & $G_H$ &
\# & \hspace{25pt} $3a_{1B}$ & \hspace{25pt} $3a_{3B}$ & $G_H$ \\
\hline \hline \endhead
1 & (-2,0,0,0,0,0,0,0) & (0,1,-1,0,0,0,0,0) & 1 & 2 & (0,2,0,0,0,0,0,0) & (-1,0,-1,0,0,0,0,0) & 1 \\ \hline
3 & (0,2,0,0,0,0,0,0) & (1,0,1,0,0,0,0,0) & 1 & 4 & (-2,0,0,0,0,0,0,0) & (0,0,0,2,1,1,1,-1) & 2 \\ \hline
5 & (-2,0,0,0,0,0,0,0) & (0,0,0,2,1,1,1,1) & 2 & 6 & (-1,1,0,1,1,0,0,0) & (-1,-1,0,0,0,0,0,0) & 2 \\ \hline
7 & (-1,1,0,1,1,0,0,0) & (0,0,0,1,-1,0,0,0) & 2 & 8 & (-1,1,0,1,1,0,0,0) & (1,1,0,0,0,0,0,0) & 2 \\ \hline
9 & (0,1,1,1,1,0,0,0) & (0,0,0,1,-1,0,0,0) & 2 & 10 & (0,1,1,1,1,0,0,0) & (0,1,-1,0,0,0,0,0) & 2 \\ \hline
11 & (0,2,0,0,0,0,0,0) & (0,0,0,2,1,1,1,-1) & 2 & 12 & (0,2,0,0,0,0,0,0) & (0,0,0,2,1,1,1,1) & 2 \\ \hline
13 & (-2,0,0,0,0,0,0,0) & (0,0,0,1,1,0,0,0) & 3 & 14 & (-1,1,0,1,1,0,0,0) & (-2,-1,-1,1,1,0,0,0) & 3 \\ \hline
15 & (-1,1,0,1,1,0,0,0) & (2,1,1,-1,-1,0,0,0) & 3 & 16 & (0,1,1,1,1,0,0,0) & (-2,-1,-1,1,1,0,0,0) & 3 \\ \hline
17 & (0,1,1,1,1,0,0,0) & (2,1,1,-1,-1,0,0,0) & 3 & 18 & (0,2,0,0,0,0,0,0) & (0,0,0,1,1,0,0,0) & 3 \\ \hline
19 & (-2,0,0,0,0,0,0,0) & (0,-1,-2,1,1,1,0,0) & 4 & 20 & (-1,1,0,1,1,0,0,0) & (-2,-1,-1,0,-1,1,0,0) & 4 \\ \hline
21 & (-1,1,0,1,1,0,0,0) & (-2,0,1,-1,-1,1,0,0) & 4 & 22 & (-1,1,0,1,1,0,0,0) & (-2,1,0,0,0,1,1,-1) & 4 \\ \hline
23 & (-1,1,0,1,1,0,0,0) & (-2,1,0,0,0,1,1,1) & 4 & 24 & (-1,1,0,1,1,0,0,0) & (2,1,1,1,0,1,0,0) & 4 \\ \hline
25 & (-1,1,0,1,1,0,0,0) & (2,-1,0,0,0,1,1,1) & 4 & 26 & (-1,1,0,1,1,0,0,0) & (-1,1,1,-1,-1,1,1,-1) & 4 \\ \hline
27 & (0,1,1,1,1,0,0,0) & (-2,-1,-1,0,-1,1,0,0) & 4 & 28 & (0,1,1,1,1,0,0,0) & (-2,1,0,1,1,1,0,0) & 4 \\ \hline
29 & (0,1,1,1,1,0,0,0) & (2,0,-1,-1,-1,1,0,0) & 4 & 30 & (0,1,1,1,1,0,0,0) & (2,1,1,1,0,1,0,0) & 4 \\ \hline
31 & (0,1,1,1,1,0,0,0) & (-1,1,1,-1,-1,1,1,-1) & 4 & 32 & (0,2,0,0,0,0,0,0) & (-2,0,1,1,1,1,0,0) & 4 \\ \hline
33 & (0,2,0,0,0,0,0,0) & (2,0,-1,1,1,1,0,0) & 4 &  & & & \\ \hline
\end{longtable}

\begin{longtable}{|r|l|l|r||r|l|l|r|}
\caption{CMM 2, $3V_{B}$ = (2,1,1,0,0,0,0,0).\label{tab2}} \\
\hline
\# & \hspace{25pt} $3a_{1B}$ & \hspace{25pt} $3a_{3B}$ & $G_H$ &
\# & \hspace{25pt} $3a_{1B}$ & \hspace{25pt} $3a_{3B}$ & $G_H$ \\
\hline \hline \endfirsthead
\caption{(continued) CMM 2, $3V_{B}$ = (2,1,1,0,0,0,0,0).} \\
\hline
\# & \hspace{25pt} $3a_{1B}$ & \hspace{25pt} $3a_{3B}$ & $G_H$ &
\# & \hspace{25pt} $3a_{1B}$ & \hspace{25pt} $3a_{3B}$ & $G_H$ \\
\hline \hline \endhead
1 & (-2,0,-1,1,0,0,0,0) & (-1,0,-1,0,0,0,0,0) & 1 & 2 & (-2,0,-1,1,0,0,0,0) & (1,0,1,0,0,0,0,0) & 1 \\ \hline
3 & (-2,1,1,0,0,0,0,0) & (0,1,-1,0,0,0,0,0) & 1 & 4 & (-2,0,-1,1,0,0,0,0) & (-2,-1,-1,1,1,0,0,0) & 2 \\ \hline
5 & (-2,0,-1,1,0,0,0,0) & (2,1,1,-1,1,0,0,0) & 2 & 6 & (-2,1,1,0,0,0,0,0) & (0,0,0,2,1,1,1,-1) & 2 \\ \hline
7 & (-2,1,1,0,0,0,0,0) & (0,0,0,2,1,1,1,1) & 2 & 8 & (-1,0,0,1,1,1,1,-1) & (-1,1,1,1,1,1,1,-1) & 2 \\ \hline
9 & (-1,0,0,1,1,1,1,-1) & (0,1,-1,0,0,0,0,0) & 2 & 10 & (-1,1,-1,1,1,1,0,0) & (-1,-1,0,0,0,0,0,0) & 2 \\ \hline
11 & (-1,1,-1,1,1,1,0,0) & (0,0,0,0,0,0,1,1) & 2 & 12 & (-1,1,-1,1,1,1,0,0) & (1,1,0,0,0,0,0,0) & 2 \\ \hline
13 & (-2,0,-1,1,0,0,0,0) & (-1,1,1,-1,1,1,1,1) & 3 & 14 & (-2,1,1,0,0,0,0,0) & (0,0,0,1,1,0,0,0) & 3 \\ \hline
15 & (-1,0,0,1,1,1,1,-1) & (0,0,0,2,1,-1,-1,1) & 3 & 16 & (-1,0,0,1,1,1,1,-1) & (0,0,0,1,1,-1,-2,-1) & 3 \\ \hline
17 & (-1,1,-1,1,1,1,0,0) & (-2,-1,-1,0,-1,-1,0,0) & 3 & 18 & (-1,1,-1,1,1,1,0,0) & (2,1,1,1,1,0,0,0) & 3 \\ \hline
19 & (-2,0,-1,1,0,0,0,0) & (-2,0,1,0,1,1,1,0) & 4 & 20 & (-2,0,-1,1,0,0,0,0) & (-2,1,0,-1,1,1,0,0) & 4 \\ \hline
21 & (-2,1,1,0,0,0,0,0) & (0,-1,-2,1,1,1,0,0) & 4 & 22 & (-1,0,0,1,1,1,1,-1) & (-2,-1,-1,0,0,-1,-1,0) & 4 \\ \hline
23 & (-1,0,0,1,1,1,1,-1) & (2,1,1,1,0,0,0,-1) & 4 & 24 & (-1,0,0,1,1,1,1,-1) & (0,-1,-2,0,-1,-1,-1,0) & 4 \\ \hline
25 & (-1,0,0,1,1,1,1,-1) & (0,-1,-2,1,1,0,0,-1) & 4 & 26 & (-1,0,0,1,1,1,1,-1) & (0,0,0,0,0,0,-1,-1) & 4 \\ \hline
27 & (-1,1,-1,1,1,1,0,0) & (-2,-1,-1,1,0,0,1,0) & 4 & 28 & (-1,1,-1,1,1,1,0,0) & (-2,0,1,0,0,-1,1,-1) & 4 \\ \hline
29 & (-1,1,-1,1,1,1,0,0) & (-2,0,1,1,-1,-1,0,0) & 4 & 30 & (-1,1,-1,1,1,1,0,0) & (-2,1,0,-1,-1,-1,0,0) & 4 \\ \hline
31 & (-1,1,-1,1,1,1,0,0) & (-2,1,0,1,1,1,0,0) & 4 & 32 & (-1,1,-1,1,1,1,0,0) & (2,1,1,0,0,-1,1,0) & 4 \\ \hline
33 & (-1,1,-1,1,1,1,0,0) & (-1,1,1,1,-1,-1,1,-1) & 4 &  & & & \\ \hline
\end{longtable}

\begin{longtable}{|r|l|l|r||r|l|l|r|}
\caption{CMM 4, $3V_{B}$ = (2,1,1,0,0,0,0,0).\label{tab4}} \\
\hline
\# & \hspace{25pt} $3a_{1B}$ & \hspace{25pt} $3a_{3B}$ & $G_H$ &
\# & \hspace{25pt} $3a_{1B}$ & \hspace{25pt} $3a_{3B}$ & $G_H$ \\
\hline \hline \endfirsthead
\caption{(continued) CMM 4, $3V_{B}$ = (2,1,1,0,0,0,0,0).} \\
\hline
\# & \hspace{25pt} $3a_{1B}$ & \hspace{25pt} $3a_{3B}$ & $G_H$ &
\# & \hspace{25pt} $3a_{1B}$ & \hspace{25pt} $3a_{3B}$ & $G_H$ \\
\hline \hline \endhead
1 & (-2,1,-1,0,0,0,0,0) & (-1,-1,0,0,0,0,0,0) & 1 & 2 & (-2,1,-1,0,0,0,0,0) & (0,-1,1,0,0,0,0,0) & 1 \\ \hline
3 & (2,-1,-1,0,0,0,0,0) & (1,1,0,0,0,0,0,0) & 1 & 4 & (-2,0,0,1,1,0,0,0) & (-2,-1,-1,1,-1,0,0,0) & 2 \\ \hline
5 & (-2,0,0,1,1,0,0,0) & (1,1,0,0,0,0,0,0) & 2 & 6 & (-2,1,-1,0,0,0,0,0) & (1,-1,-1,1,1,1,1,-1) & 2 \\ \hline
7 & (-2,1,-1,0,0,0,0,0) & (1,-1,-1,1,1,1,1,1) & 2 & 8 & (-1,1,0,1,1,1,1,0) & (-1,0,-1,0,0,0,0,0) & 2 \\ \hline
9 & (-1,1,0,1,1,1,1,0) & (-1,1,1,-1,-1,-1,-1,1) & 2 & 10 & (-1,1,0,1,1,1,1,0) & (0,1,-1,0,0,0,0,0) & 2 \\ \hline
11 & (2,-1,-1,0,0,0,0,0) & (1,-1,-1,1,1,1,1,-1) & 2 & 12 & (2,-1,-1,0,0,0,0,0) & (1,-1,-1,1,1,1,1,1) & 2 \\ \hline
13 & (-2,0,0,1,1,0,0,0) & (2,1,1,1,1,0,0,0) & 3 & 14 & (-2,0,0,1,1,0,0,0) & (-1,1,1,1,1,1,1,-1) & 3 \\ \hline
15 & (-2,1,-1,0,0,0,0,0) & (-2,-1,-1,1,1,0,0,0) & 3 & 16 & (-1,1,0,1,1,1,1,0) & (2,1,1,0,0,0,-1,-1) & 3 \\ \hline
17 & (-1,1,0,1,1,1,1,0) & (-1,1,1,-1,-1,-1,-1,-1) & 3 & 18 & (2,-1,-1,0,0,0,0,0) & (-2,-1,-1,1,1,0,0,0) & 3 \\ \hline
19 & (-2,0,0,1,1,0,0,0) & (-2,-1,-1,0,0,1,1,0) & 4 & 20 & (-2,0,0,1,1,0,0,0) & (-2,1,0,0,0,1,1,-1) & 4 \\ \hline
21 & (-2,0,0,1,1,0,0,0) & (-2,1,0,0,0,1,1,1) & 4 & 22 & (-2,0,0,1,1,0,0,0) & (2,1,1,0,-1,1,0,0) & 4 \\ \hline
23 & (-2,0,0,1,1,0,0,0) & (0,0,0,1,0,1,0,0) & 4 & 24 & (-2,1,-1,0,0,0,0,0) & (2,-1,0,1,1,1,0,0) & 4 \\ \hline
25 & (-2,1,-1,0,0,0,0,0) & (0,-1,-2,1,1,1,0,0) & 4 & 26 & (-1,1,0,1,1,1,1,0) & (-2,-1,-1,1,0,0,-1,0) & 4 \\ \hline
27 & (-1,1,0,1,1,1,1,0) & (-2,0,1,1,1,0,0,-1) & 4 & 28 & (-1,1,0,1,1,1,1,0) & (-2,1,0,0,0,-1,-1,1) & 4 \\ \hline
29 & (-1,1,0,1,1,1,1,0) & (2,1,1,0,0,0,-1,1) & 4 & 30 & (-1,1,0,1,1,1,1,0) & (2,-1,0,0,0,-1,-1,1) & 4 \\ \hline
31 & (-1,1,0,1,1,1,1,0) & (2,0,-1,0,-1,-1,-1,0) & 4 & 32 & (-1,1,0,1,1,1,1,0) & (-1,1,1,1,1,1,-1,1) & 4 \\ \hline
33 & (2,-1,-1,0,0,0,0,0) & (-2,1,0,1,1,1,0,0) & 4 &  & & & \\ \hline
\end{longtable}

\begin{longtable}{|r|l|l|r||r|l|l|r|}
\caption{CMM 6, $3V_{B}$ = (2,1,1,0,0,0,0,0).\label{tab6}} \\
\hline
\# & \hspace{25pt} $3a_{1B}$ & \hspace{25pt} $3a_{3B}$ & $G_H$ &
\# & \hspace{25pt} $3a_{1B}$ & \hspace{25pt} $3a_{3B}$ & $G_H$ \\
\hline \hline \endfirsthead
\caption{(continued) CMM 6, $3V_{B}$ = (2,1,1,0,0,0,0,0).} \\
\hline
\# & \hspace{25pt} $3a_{1B}$ & \hspace{25pt} $3a_{3B}$ & $G_H$ &
\# & \hspace{25pt} $3a_{1B}$ & \hspace{25pt} $3a_{3B}$ & $G_H$ \\
\hline \hline \endhead
1 & (-1,0,0,1,0,0,0,0) & (-2,0,-1,1,0,0,0,0) & 1 & 2 & (0,1,0,1,0,0,0,0) & (-2,-1,0,1,0,0,0,0) & 1 \\ \hline
3 & (0,1,0,1,0,0,0,0) & (0,-1,2,1,0,0,0,0) & 1 & 4 & (-1,0,0,1,0,0,0,0) & (-1,0,0,-1,1,1,1,-1) & 2 \\ \hline
5 & (-1,0,0,1,0,0,0,0) & (2,0,0,-1,1,0,0,0) & 2 & 6 & (0,1,0,1,0,0,0,0) & (-2,0,-1,0,1,0,0,0) & 2 \\ \hline
7 & (0,1,0,1,0,0,0,0) & (0,-2,0,-1,1,0,0,0) & 2 & 8 & (-2,0,-1,1,1,1,0,0) & (-1,1,-1,1,1,1,0,0) & 2 \\ \hline
9 & (-2,0,-1,1,1,1,0,0) & (0,1,0,1,1,1,1,1) & 2 & 10 & (-2,0,-1,1,1,1,0,0) & (1,1,1,1,1,1,0,0) & 2 \\ \hline
11 & (-2,1,1,1,1,0,0,0) & (0,2,-1,0,-1,0,0,0) & 2 & 12 & (-2,1,1,1,1,0,0,0) & (0,0,-2,1,1,0,0,0) & 2 \\ \hline
13 & (-1,0,0,1,0,0,0,0) & (-1,0,0,-1,1,1,1,1) & 3 & 14 & (0,1,0,1,0,0,0,0) & (-1,-1,1,1,1,1,0,0) & 3 \\ \hline
15 & (-2,0,-1,1,1,1,0,0) & (-2,0,-1,1,0,0,0,0) & 3 & 16 & (-2,0,-1,1,1,1,0,0) & (-1,-1,1,1,-1,-1,0,0) & 3 \\ \hline
17 & (-2,1,1,1,1,0,0,0) & (-2,1,1,0,0,0,0,0) & 3 & 18 & (-2,1,1,1,1,0,0,0) & (2,0,0,-1,-1,0,0,0) & 3 \\ \hline
19 & (-1,0,0,1,0,0,0,0) & (-1,1,-1,-1,1,1,0,0) & 4 & 20 & (0,1,0,1,0,0,0,0) & (2,0,0,0,1,1,0,0) & 4 \\ \hline
21 & (0,1,0,1,0,0,0,0) & (-1,1,-1,-1,1,1,0,0) & 4 & 22 & (-2,0,-1,1,1,1,0,0) & (-2,-1,0,0,0,-1,0,0) & 4 \\ \hline
23 & (-2,0,-1,1,1,1,0,0) & (-2,1,1,0,0,0,0,0) & 4 & 24 & (-2,0,-1,1,1,1,0,0) & (-1,-1,1,0,0,-1,1,-1) & 4 \\ \hline
25 & (-2,0,-1,1,1,1,0,0) & (-1,-1,1,0,0,-1,1,1) & 4 & 26 & (-2,0,-1,1,1,1,0,0) & (-1,-1,1,1,1,0,1,0) & 4 \\ \hline
27 & (-2,0,-1,1,1,1,0,0) & (2,0,0,0,-1,-1,0,0) & 4 & 28 & (-2,0,-1,1,1,1,0,0) & (-1,1,-1,-1,-1,-1,0,0) & 4 \\ \hline
29 & (-2,1,1,1,1,0,0,0) & (-2,0,-1,0,0,1,0,0) & 4 & 30 & (-2,1,1,1,1,0,0,0) & (-1,0,0,-1,-1,1,1,-1) & 4 \\ \hline
31 & (-2,1,1,1,1,0,0,0) & (2,0,0,1,0,1,0,0) & 4 & 32 & (-2,1,1,1,1,0,0,0) & (-1,1,-1,-1,-1,1,0,0) & 4 \\ \hline
33 & (-2,1,1,1,1,0,0,0) & (0,-1,-1,0,-1,1,1,1) & 4 &  & & & \\ \hline
\end{longtable}

\begin{longtable}{|r|l|l|r||r|l|l|r|}
\caption{CMM 8, $3V_{B}$ = (1,1,0,0,0,0,0,0).\label{tab8a}} \\
\hline
\# & \hspace{25pt} $3a_{1B}$ & \hspace{25pt} $3a_{3B}$ & $G_H$ &
\# & \hspace{25pt} $3a_{1B}$ & \hspace{25pt} $3a_{3B}$ & $G_H$ \\
\hline \hline \endfirsthead
\caption{(continued) CMM 8, $3V_{B}$ = (1,1,0,0,0,0,0,0).} \\
\hline
\# & \hspace{25pt} $3a_{1B}$ & \hspace{25pt} $3a_{3B}$ & $G_H$ &
\# & \hspace{25pt} $3a_{1B}$ & \hspace{25pt} $3a_{3B}$ & $G_H$ \\
\hline \hline \endhead
1 & (0,0,1,1,1,1,0,0) & (0,0,-1,-1,-1,-1,1,1) & 1 & 2 & (0,0,1,1,1,1,0,0) & (-1,-2,0,0,0,-1,0,0) & 2 \\ \hline
3 & (0,0,1,1,1,1,0,0) & (0,0,2,0,0,0,1,-1) & 2 & 4 & (0,0,1,1,1,1,0,0) & (0,0,-1,-1,-1,-1,1,-1) & 3 \\ \hline
5 & (0,0,1,1,1,1,0,0) & (0,0,2,0,0,0,1,1) & 4 &  & & & \\ \hline
\end{longtable}

\begin{longtable}{|r|l|l|r||r|l|l|r|}
\caption{CMM 8, $3V_{B}$ = (2,1,1,1,1,0,0,0).\label{tab8b}} \\
\hline
\# & \hspace{25pt} $3a_{1B}$ & \hspace{25pt} $3a_{3B}$ & $G_H$ &
\# & \hspace{25pt} $3a_{1B}$ & \hspace{25pt} $3a_{3B}$ & $G_H$ \\
\hline \hline \endfirsthead
\caption{(continued) CMM 8, $3V_{B}$ = (2,1,1,1,1,0,0,0).} \\
\hline
\# & \hspace{25pt} $3a_{1B}$ & \hspace{25pt} $3a_{3B}$ & $G_H$ &
\# & \hspace{25pt} $3a_{1B}$ & \hspace{25pt} $3a_{3B}$ & $G_H$ \\
\hline \hline \endhead
1 & (-1,0,0,0,-1,1,1,0) & (-1,-1,-1,-1,-1,0,0,-1) & 2 & 2 & (-1,0,0,0,-1,1,1,0) & (-1,1,1,1,-1,0,0,1) & 2 \\ \hline
3 & (0,1,1,1,0,1,0,0) & (1,-1,-1,-1,1,-1,0,0) & 2 & 4 & (-1,0,0,0,-1,1,1,0) & (-1,1,1,0,0,-1,-1,1) & 3 \\ \hline
5 & (-1,0,0,0,-1,1,1,0) & (0,0,-1,-1,-1,-1,-1,-1) & 3 & 6 & (-1,1,1,0,0,1,0,0) & (-2,0,0,-1,-1,0,0,0) & 3 \\ \hline
7 & (-1,1,1,0,0,1,0,0) & (1,1,1,1,1,1,0,0) & 3 & 8 & (-1,0,0,0,-1,1,1,0) & (-2,0,-1,-1,0,0,0,0) & 4 \\ \hline
9 & (-1,0,0,0,-1,1,1,0) & (-2,0,0,0,1,1,0,0) & 4 & 10 & (-1,0,0,0,-1,1,1,0) & (-2,1,0,0,0,0,0,1) & 4 \\ \hline
11 & (-1,0,0,0,-1,1,1,0) & (-1,0,-1,-1,1,0,-1,1) & 4 & 12 & (-1,0,0,0,-1,1,1,0) & (-1,1,1,-1,1,0,-1,0) & 4 \\ \hline
13 & (-1,1,1,0,0,1,0,0) & (-2,1,0,0,0,-1,0,0) & 4 & 14 & (-1,1,1,0,0,1,0,0) & (-1,-1,-1,-1,-1,0,1,0) & 4 \\ \hline
15 & (-1,1,1,0,0,1,0,0) & (-1,-1,-1,1,0,0,1,-1) & 4 & 16 & (0,1,1,1,0,1,0,0) & (-2,0,0,0,1,-1,0,0) & 4 \\ \hline
17 & (0,1,1,1,0,1,0,0) & (0,1,0,-2,1,0,0,0) & 4 &  & & & \\ \hline
\end{longtable}

\begin{longtable}{|r|l|l|r||r|l|l|r|}
\caption{CMM 9, $3V_{B}$ = (1,1,0,0,0,0,0,0).\label{tab9a}} \\
\hline
\# & \hspace{25pt} $3a_{1B}$ & \hspace{25pt} $3a_{3B}$ & $G_H$ &
\# & \hspace{25pt} $3a_{1B}$ & \hspace{25pt} $3a_{3B}$ & $G_H$ \\
\hline \hline \endfirsthead
\caption{(continued) CMM 9, $3V_{B}$ = (1,1,0,0,0,0,0,0).} \\
\hline
\# & \hspace{25pt} $3a_{1B}$ & \hspace{25pt} $3a_{3B}$ & $G_H$ &
\# & \hspace{25pt} $3a_{1B}$ & \hspace{25pt} $3a_{3B}$ & $G_H$ \\
\hline \hline \endhead
1 & (1,0,1,0,0,0,0,0) & (-1,1,-1,1,1,1,1,1) & 1 & 2 & (1,0,1,0,0,0,0,0) & (1,2,0,1,1,1,0,0) & 2 \\ \hline
3 & (-1,-1,1,1,1,1,1,-1) & (0,0,2,0,-1,-1,-1,1) & 2 & 4 & (-1,-1,1,1,1,1,1,-1) & (0,0,0,0,0,-1,-1,0) & 3 \\ \hline
5 & (-1,-1,1,1,1,1,1,-1) & (-1,-2,1,0,0,0,-1,-1) & 4 &  & & & \\ \hline
\end{longtable}

\begin{longtable}{|r|l|l|r||r|l|l|r|}
\caption{CMM 9, $3V_{B}$ = (2,1,1,1,1,0,0,0).\label{tab9b}} \\
\hline
\# & \hspace{25pt} $3a_{1B}$ & \hspace{25pt} $3a_{3B}$ & $G_H$ &
\# & \hspace{25pt} $3a_{1B}$ & \hspace{25pt} $3a_{3B}$ & $G_H$ \\
\hline \hline \endfirsthead
\caption{(continued) CMM 9, $3V_{B}$ = (2,1,1,1,1,0,0,0).} \\
\hline
\# & \hspace{25pt} $3a_{1B}$ & \hspace{25pt} $3a_{3B}$ & $G_H$ &
\# & \hspace{25pt} $3a_{1B}$ & \hspace{25pt} $3a_{3B}$ & $G_H$ \\
\hline \hline \endhead
1 & (-1,0,0,0,0,1,0,0) & (-1,0,0,0,-1,0,0,0) & 2 & 2 & (0,1,0,0,0,1,0,0) & (-1,-1,-1,-1,-1,-1,1,1) & 2 \\ \hline
3 & (-2,0,0,0,-1,1,1,1) & (-1,-1,-1,-1,-1,1,1,-1) & 2 & 4 & (-1,0,0,0,0,1,0,0) & (-2,1,0,0,0,-1,1,-1) & 3 \\ \hline
5 & (-2,0,0,0,-1,1,1,1) & (-2,0,0,-1,-1,1,1,0) & 3 & 6 & (-1,0,0,0,0,1,0,0) & (-2,0,0,-1,-1,-1,1,0) & 4 \\ \hline
7 & (0,1,0,0,0,1,0,0) & (-2,-1,0,0,-1,-1,1,0) & 4 & 8 & (-2,0,0,0,-1,1,1,-1) & (-2,0,0,-1,-1,1,0,-1) & 4 \\ \hline
9 & (-2,0,0,0,-1,1,1,-1) & (-2,1,0,0,0,-1,-1,1) & 4 & 10 & (-2,0,0,0,-1,1,1,-1) & (-2,1,0,0,0,1,1,-1) & 4 \\ \hline
11 & (-2,0,0,0,-1,1,1,-1) & (-1,-1,-1,-1,-1,1,-1,-1) & 4 & 12 & (-2,0,0,0,-1,1,1,-1) & (-1,1,1,-1,1,-1,-1,1) & 4 \\ \hline
13 & (-2,0,0,0,-1,1,1,-1) & (-1,1,1,1,-1,1,-1,-1) & 4 & 14 & (-2,0,0,0,-1,1,1,-1) & (1,-1,-1,-1,1,1,-1,-1) & 4 \\ \hline
15 & (-2,0,0,0,-1,1,1,-1) & (1,1,1,1,1,1,-1,-1) & 4 & 16 & (-2,0,0,0,-1,1,1,1) & (-2,0,-1,-1,0,1,0,-1) & 4 \\ \hline
\end{longtable}
}

\end{document}